\documentclass[]{aa}
\usepackage{psfig}
\begin{document}

\title{Spectrophotometry and structural analysis of 5 comets
\thanks{Based
on observations taken at the German-Spanish
Astronomical Centre, Calar Alto, operated by the Max-Planck-Institute for
Astronomy, Heidelberg, jointly with the Spanish National Commission
for Astronomy}}

\titlerunning{Spectrophotometry and morphology of 5 comets}
\authorrunning{Szab\'o M. Gy. et al.}
\offprints{l.kiss@physx.u-szeged.hu}

\author{
Gy. M. Szab\'o\inst{1} \and
L. L. Kiss\inst{1} \and
K. S\'arneczky\inst{2} \and
K. Szil\'adi\inst{1}
}

\institute{
Department of Experimental Physics \& Astronomical Observatory,
University of Szeged,
H-6720 Szeged, D\'om t\'er 9., Hungary \and
 Department of Physical Geography, ELTE University, H-1088 Budapest,
Ludovika t\'er 2., Hungary}

\date{}

\abstract{
We discuss the morphology and spectrophotometry of 5 comets visible
in August, 2001.
We decompose comae into coma profiles and azimuthally renormalized
images, in which general and local features are quantitatively
comparable.
Comet 19P/Borrelly showed a strong gas fan toward the solar direction,
but no detectable gas in the tail.
Dust in its inner coma was collimated toward the antisolar
direction and the tail, with no dust in the outer coma.
The contribution of spatial variations structure was moderate, about 35\%.
Comet 29P Schwassmann-Wachmann 1
was observed in outburst: we detected ``spinning'' jet structures.
A high level of dust production resulted in an unusually
high $Af{\rho}$=16600 cm.
The spatial variations reached $-$77\%,
at the minimum, due in part to a jet and
a ring-like structure in 1 arcminute distance from the nucleus.
In comet C/2001 A2,
we detected a strong post-perihelion increase of dust and gas activity,
in which the C$_2$ profile became one magnitude brighter over
a 3-day period.
For comets C/2000 SV74 and C/2000 WM1,
we present detailed pre-perihelion spectrophotometry 
and morphological information.
Comet C/2000 SV74 showed
high dust production ($Af{\rho}=1479$ cm).
Its coma suggests
a steady-state outflow of material,
while the low contribution of spatial variations support
high level activity.
The coma of C/2000 WM1 is dominated by solar effects,
and CO+ forms the bulk of its gas activity.
Despite its large heliocentric distance, we observed a nice tail.
}

\maketitle

\section{Introduction}

The distribution of distinctive cometary components (dust, ions, radicals)
can be well studied with help of two-dimensional, narrow-band CCD images. 
In addition to measuring the spatial distribution of species,
one can measure column densities of gas and
dust, and draw conclusions on chemical composition, 
production rates and activity levels. 
Despite the efficiency of this method, only a small
fraction of visible comets are studied with detailed spectrophotometry. 
Spectroscopic methods, on the other hand,
have the advantage of producing a spectrum with a 
well-defined continuum level, so emission features show clearly.
Large spectrophotometric and spectroscopic
surveys of many comets were published by A'Hearn et al.
(1995) (hereafter A95) and Fink \& Hicks (1996), where the interested reader
can find detailed descriptions of methods.

Surface photometry of images can address coma morphology
(such as radial coma profiles and non-radial features, also called
as coma profiles and azimuthally renormalized images, see, e.g., Lederer et
al. 1997, Larson \& Slaughter 1991),
and also may yield estimated nuclear radii (e.g. Luu \& Jewitt 1992, Lamy
\& T\'oth 1995, Lowry et al. 1999). 
An appropriate selection of
medium- and narrow-band filters
centered on different wavelengths 
can separate the dust continuum from emission by gas.
Differences in
gas and dust components reveal the effect of radiation
pressure on different types of particles (a good example of
combined quantitative coma analysis can be found, e.g. in Schulz et al. 1993).

The main aim of our work is to contribute to this field of ground-based
solar system research with new narrow- and medium-band
spectrophotometric observations of comets visible in August, 2001. The
function of this paper is to present the results of observations 
of 5 comets carried out at Calar Alto Observatory. The paper is organised 
as follows. Sect.\ 2 deals with the observations, methods of analysis are
described in Sect.\ 3, results on individual objects are given in Sect.\ 4,
while the discussion is presented in Sect.\ 5. This work is an extension
of our previous analysis of distant active comets (Szab\'o et al. 2001).

\section{Observing strategy and data acquisition}

%Table 1.
\begin{table}
\caption{Filter characteristics. Band is shown in nm/nm, filter diameter D in 
mm}
\label{Filter}
\begin{center}
\begin{tabular} {lllll}
\hline   
Filter & Band & Transm. & D(mm) & Code \\
\hline
Comet CN & 387.1/5.0  & 35\% &42& CN \\
Gunn v   & 411.1/37.4 & 19\% &50& v\\
Comet CO+& 425.4/6.9  & 56\% &32& CO+\\
Gunn g   & 492.5/54.0 & 74\% &50& g\\
Comet C$_2$ & 512.5/12.5 & 65\% &42& C$_2$\\
Gunn r   & 662.6/104.5& 77\% &50& r\\
753/30   & 751.5/28.2 & 57\% &50& j\\ 
Gunn z   & 910.0/90.0 & 90\% &50& z\\
\hline
\end{tabular}
\end{center}
\end{table}

%Fig. 1.
\begin{figure}
\begin{center}
\leavevmode
\caption{Filter curves of
our filters compared to a model spectrum of a comet showing
most of the characteristic emission features.
Italic labelled solid lines show the Gunn filters (simplified curves) and the
753/30 interference filter (referred as j in this paper), while
dashed lines pertain to our special comet filters.
}
\end{center}
\label{f1}
\end{figure}

Our primary goal was to obtain spectrophotometric observations that
would enable morphological as well as quantitative studies of bright and
moderately bright comets. The most important constraints were: 1) to
separate comet continuum and emission bands and 2) to record the detected
signal at the high S/N level required for detailed morphological
calculations. 
The first required that the selected photometric system could
exclude strong emission bands, while the second point required the use of
medium- or wide-band photometry.

When selecting the filters,
the main aim was to use medium-band filters
to measure the continuum in the optical range up to 1$\mu$.
The wide-band Johnson BVRI system is not well-suited for this kind of
work because it includes strong emission features in cometary
spectra. 
Furthermore, the overlaps between the Johnson filters reduce 
their ability
to trace continuum variations. 
Therefore, we decided to use the Gunn
$v$, $g$, $r$, $z$ filter sequence (Thuan \& Gunn 1976), which is used
mostly by extragalactic researchers. 
For quantitative spectrophotometry, 
we added interference filters: CN, CO+ and C$_2$
(see Table\ 1). 
As the wide-band standard Gunn $i$ filter (Wade et al. 1979)
passes the band of CN (2--0), we replaced it by an interference filter
753/30, similar in effective wavelength, which refer to as $j$ in the
Tables. 
The Gunn $z$ filter includes the 1
$\mu$m border of the continuum, and it also passes the CN (1--0)
resonance line. 
Two of the interference filters, CN and CO+, just border
the band of Gunn $v$. 
The Gunn $g$ filter includes 70 percent of the C$_2$ filter
band, which must be taken into account in reductions.
Gunn $r$ includes the
NH$_2$(0,10,0) band, 
which is not too strong even in the most active comets and
does not influence the continuum flux significantly. 
A further advantage is
that these filters avoid emission lines of airglow and light pollution,
and so the background is significantly darker than it would be with
Johnson filters. 
Table\ 1 summarizes the filter characteristics as
given by the manufacturer.
In Figure\ 1, we compare our passbands to a typical model comet spectrum 
(Lowry et al. 1999, Arpigny 1995)

The observations were obtained at Calar Alto Observatory, with the 1.23
meter telescope equipped with a SITE\#2b CCD camera on five nights
in August, 2001. The image scale of the
unbinned frames is 0.49 \arcsec/pixel. 
In order to avoid effects of image
trailing, the telescope was set to follow the motion of the comets, not
the ordinary sidereal rate. 
We chose
comets predicted to be brighter than $r$=15\fm0
(19P/Borrelly, 29P/Schwassmann-Wachmann 1, C/2000 SV74 LINEAR, C/2000 WM1
LINEAR, C/2001 A2 LINEAR) as targets. Their aspect data are
summarized in Table 2. The full observing log is presented in Table\ 3. 

The atmospheric extinction in the Gunn-system was monitored by repeated
observations of BD+17$^\circ$4708 selected from the lists of Thuan \&
Gunn (1976). The standard photometric transformations were determined 
by observations of Gunn-standards in the field of M~34 (Kent 1985) 
several times per night. Typical photometric uncertainties were 
$\pm$0\fm015 in the extinction determination and $\pm$0\fm015 in the
zero-points of standard transformations. 
The overall uncertainty is
estimated to be $\pm$0\fm022-0\fm033, depending on the wavelength and
target brightness. Two additional spectrophotometric standard stars
(HD~187811 and HD~183439A) selected from Alekseeva et al.
(1996) were also observed on Aug. 13, 2001 (MJD 52135.4).

%Table 2.
\begin{table*}
\caption{Aspect data of comets referred to the middle of the observing run.
Signs used are: MJD modified Julian-Date, $\Delta$ geocentric radius, $R$
heliocentric radius, $\epsilon$ elongation, $\alpha$ solar phase, $\Sigma$
solar direction position angle $\mu_{\rm r}$ sky brightness through Gunn r
filter, See. seeing. MJD 52134.5 = 13th August, 2001, 0:00 UT.}
\begin{center}
\begin{tabular} {llllrrrllllr}
\hline
Comet & MJD & $\Delta$(AU) & $R$(AU) & $\epsilon$ & $\alpha$ & $\Sigma$ &
$\mu_{\rm r}$ & See. & Airmass & Scale (km/\arcsec) & Code \\
\hline
19P/Borrelly& 52137.66& 1.685 & 1.400 & 56.8  & 36.4 &87.6 & 20.3 & 2.2 & 2.9--1.9 &
1225 &19P \\
29P/SW1     & 52135.35 & 5.121 & 5.919 & 138.6 & 6.5  &294.8 & 20.4 & 1.4 & 2.4--2.6 &
3724 &29P \\
C/2000 SV74&52135.53 & 4.036 & 4.244 & 93.3  & 13.8 &102.0 & 20.7 & 1.0 & 1.6--1.4 &
2934& SV74 \\
C/2000 WM1& 52137.55 & 2.891 & 2.789 & 74.1  & 20.4 &115.4 & 20.9 & 1.1 & 2.0--1.7 &
2102 & WM1 \\ 
C/2001 A2 & 52134.51 & 0.714 & 1.628 & 140.4 & 23.4 &357.7 & 20 & 1.0 & 1.1 &
519 & A2a\\
C/2001 A2 & 52136.50 & 0.747 & 1.655 & 139.8 & 23.2 &357.6 & 20 & 1.1 & 1.4 &
543 & A2b\\
C/2001 A2 & 52137.52 & 0.764 & 1.669 & 139.5 & 23.2 &357.6 & 20.9 & 1.1 & 1.1 &
555 & A2c\\ 
\hline
\end{tabular}
\end{center}
\end{table*}

%Table 3.
\begin{table}
\caption{Log of the individual images. Exp. means integrated exposure time 
in {\it "number of images"} $\times$ {\it "exposure time of one individual"} 
format. Remarks are: (1): has not been detected, or has been not brighter 
than about one fourth of the background scatter on the individual images.
(2): Effect of clouds is visible on images of the third sequence.}
\label{Images}
\begin{center}
\begin{tabular} {llrl}
\hline   
Code & Filter & Exp.(s) & Remark \\
\hline
{\bf 19P} & v,g,j               & 3$\times$60 & \\
          & r,z,CO+,C$_2$,CN       & 6$\times$60 \\
{\bf 29P} & v,g,r,j,z,CN,CO+,C$_2$ & 3$\times$240 & \\
{\bf SV74}& v,g,r,j,z,C$_2$        & 3$\times$120 &\\
          & CO+,CN              & 1$\times$120 & (1)\\
{\bf WM1} & v,g,r,j,z,C$_2$        & 3$\times$90 & \\
          & CN,CO+              & 1$\times$90 & (1)\\
{\bf A2a} & g,r,C$_2$,CO+,CN       & 3$\times$60 & \\
{\bf A2b} & r,v,g,j,z,CN,CO+,C$_2$ & 2$\times$90 & (2)\\
{\bf A2c} & v,g,r,j,z,CN,C$_2$     & 3$\times$105 & \\
          & CO+                 & 1$\times$120 & (1)\\
\hline
\end{tabular}
\end{center}
\end{table}

\section{Analysis and data reduction}

The data obtained through the interference filters were calibrated following
the usual methods of reduction as defined by A'Hearn (1983). Briefly, field
stars on the same frames were measured with the narrow-band comet filters
to find their $m_{\rm instr.}$ instrumental magnitudes. These values
%
%WMR
%
were combined with the standard Gunn magnitudes and color indices in a
least-squares fit to yield the color coefficients and stellar zero-points
($\kappa_{\rm XX}$). For a given comet, the same coefficients and zero-points
were used to determine expected narrow-band magnitudes, assuming that
light is produced only by the comet's continuum. 
Then the difference of the observed
and expected narrow-band magnitudes is assigned to the emission bands.
The resulting equations are:

$m^{\rm CN,stellar}_{\rm instr.}=v+0\fm43(v-g)+\kappa_{\rm CN}$

$m^{\rm CO+,stellar}_{\rm instr.}=v-0\fm15(v-g)+\kappa_{\rm CO}$

$m^{\rm C_2,stellar}_{\rm instr.}=g-0\fm09(v-g)+\kappa_{\rm C_2}$.

\noindent The $\kappa$ zero points vary slightly, depending on sky conditions:
roughly
2\fm2 for CN and CO+ and 1\fm5 for C$_2$; rms scatter of the fits
to field stars is smaller than 0\fm04.

Having separated the emission and continuum,
we calculated absolute fluxes of emission
features based on observations of
spectrophotometric standards HD~183439A and HD~187811 (Alekseeva et al.
1996). We determined flux excesses $F_{\rm CN}$, $F_{\rm CO+}$, $F_{\rm C_2}$;
$F_{\rm XX}=F_{\rm XX,observed}-F_{\rm XX,stellar}$.
For calculating the total luminosity, we assumed isotropic radiation
from the coma:
$L({\rm total})=L({\rm measured}) \cdot 4 \pi \Delta ^2 (m^2)$.

To describe dust continuum production, we adopted the
$Af{\rho}$ values from A'Hearn (1983).
The $\rho$ radius was chosen
by selecting the linear part of the surface brightness profile. The
calculation also includes the flux ratio of the comet to the Sun.
We needed the Gunn $r$ magnitude of the Sun, which was calculated from the
Johnson R$_\odot=-27\fm26$ transformed to Gunn $r$ according to the formula
of J\o rgensen (1996) $r-R=0\fm354$, rms=0\fm035, resulting
$r_{\sun}=-26\fm91$. Gas production parameters describe the number of radicals
ejected from the nucleus during one second. 
The fluorescence efficiencies ($g$-factors) and scale
lengths were taken from A95 for CN and C$_2$ and from Cochran
et al. (2000) for CO+. 

To our knowledge, there are no precise Gunn colors of the Sun in the
literature. However, we needed them to study the comet continuum
reflectance, where spectral variations of the Sun have to be taken into
account. Therefore, we measured precise colors of several main-belt 
asteroids with previously published medium-resolution optical spectra 
during
our observing run. The best target was 2 Pallas, which has a
largely constant reflectance spectrum over the visible wavelength (Gaffey et
al. 1993), making it easy to transform its Gunn-colors to those of the Sun.
Without giving the details (Szab\'o et al. 2002, in prep.) we estimate the
following values: (v--g)$_{\sun}$=0.254, (g--r)$_{\sun}$=0.126,
(r--j)$_{\sun}$=0.006, (j--z)$_{\sun}$=$-$0.009, with photometric errors about
$\pm$0\fm02. 

As usual in morphological studies, field stars had to
be subtracted from the comae. This was done with the DAOPHOT
package in IRAF. For correct subtraction in the coma
the foreground coma contribution
was estimated based on manual examination of the coma,
near the individual stars. The pixel intensities were modified that the
background is zero and the integrated flux is 1.
That allowed to isolate the coma contribution
by a median combination of these star-subtracted images.

Further examinations were based on some tools commonly used in, e.g., galaxy
morphology (see eg. Ravindranath et al. 2001 for recent discussion). To
extract the surface brightness change across the nucleus, we used the {\it
apextract} task from the TWODSPEC package in IRAF. A 5\farcs0-wide aperture
was shifted through the coma across the nucleus in two sampling directions.
One of them was the solar-antisolar line (hereafter referred as radial
section), the other perpendicular to the radial direction
(hereafter referred as tangential section).  {\it Azimuthally renormalized
images} are
similar to the residual images defined by the difference of the coma image
and an analytic radial profile. But instead of an analytic fit, we simply
used the radial coma image resulting from an azimuthally averaged coma image.
``Bright'', positive areas in the residual image refer to matter excess
while ``dark'', negative values show tenuous areas. 
This method
emphasizes the presentation of special phenomena,
such as the ellipticity of the coma, jet or spin structures.
To describe these features, we extracted 
surface brightness profiles from
the residual images, just as we did for the original images. 

For physical conclusions, the strength of spatial variations must be
characterized. We calculated {\it local intensity ratios} for
the negative and positive peaks of azimuthally renormalized images. 
They are given by the
peak intensities with respect to the normalised coma intensity at the same
position. 

\section{Results}

%Table 4.
\begin{table*}
\caption{Magnitudes, colors and production parameters. 
$L$ fluxes in $10^{15}$erg~s$^{-1}$,
${\rho}$ in $10^7$ cm and $Af{\rho}$ in cm.}
\label{MagCoProd}
\begin{center}
\begin{tabular} {|llrrrrrr|}
\hline
Ref. & $r$ & $v-g$ & $g-r$ & $r-j$ & $j-z$ & $L_{\rm CN}$ & $L_{\rm C_2}$ \\
\hline
19P & 12.27 & $-$0.07 & 0.26 & $-$0.04 & 0.19 & 2150 & 531 \\
29P & 12.68 & 0.33 & $-$0.07 & 0.32 & 0.01 & $<$234&$<$593 \\ 
SV74 & 15.29 & 0.22 & 0.58   & 0.22 & 0.02    &$<$23 & 132 \\ 
WM1 & 15.19 & 0.13 & 0.28 & 0.33 & $-$0.15    &$<$70  & 63 \\ 
A2a & 13.52 & -- & $-$0.06 & -- & --          &126   & 57 \\  
A2c & 13.46 & $-$0.10 & $-$0.11 & 0.11 & 0.15 &830   & 274 \\
\hline
Ref. & $L_{\rm CO+}$ & ${\rho}$ & slope & log $Af{\rho}$ & [C$_2$]-[CN]$^a$ & [$Af{\rho}$]-[C$_2$] &[$Af{\rho}$]-[CN]\\  
\hline
19P & $<$52 & 60.5 & $-$0.97 & 2.98 & $-$0.38 & $-$22.32 &$-$22.70\\
29P & $<$329& 148.0 & $-$1.60 & 4.22 & -- &$>-$21.10 & $>-$20.50\\
SV74 & 209   & 146.0 & $-$1.21 & 3.17 & $>$0.6 & $-$21.12 & $>-$20.52\\
WM1 & 396   & 105.0 &$-$1.42 & 2.72 & $>$0.2 & $-$21.99 & $>-$21.80\\
A2a & $<$4  & 22.5 & $-$1.04 & 2.21 & $-$0.12 & $-$22.07 & $-$22.19\\
A2c & $<$4  & 27 & $-$1.07 & 2.28 & $-$0.26 & $-$22.74 & $-$23.00 \\
\hline
\end{tabular}
\end{center}
$^a$ Using log(Q(C$_2$)/Q(CN)), log($Af{\rho}$/Q(C$_2$)), log($Af{\rho}$/Q(CN)),
production rates in mol sec$^{-1}$, $Af{\rho}$ in cm.
\end{table*}

Absolute photometric data and calculated production parameters of comets are
compared in Table\ 4. Values of $r$ magnitude and color indices are measured
in apertures of 10\arcsec\ radius, as well as the $L$ total fluxes of emission
features.  They are followed by ${\rho}$ [$10^7$ cm] radii, $slope$ of the
coma profile inside ${\rho}$ radius and $Af{\rho}$ values measured inside
${\rho}$.

Remarks on the individual comets are as follows.
\bigskip\newline
\noindent
{\it 19P/Borrelly}
\newline
\noindent

This comet belongs to the Jupiter family, with an orbital period of about 7
years. It is the prototype of the Borrelly-type class of comets (defined by
Fink et al. 1999), known for its low C$_2$ production. The 1994 perihelion was
studied by A95, who determined several production rates. A detailed
structural analysis of its coma and nucleus is presented by Lamy et al.
(1998) based on observations taken during the same apparition. They have
determined a prolate spheroid nucleus model with 4.4$\times$1.8 km
semi-axes. The estimated fractional active area is 8\%. 
Our observations were made about two months before
the Deep Space-1 spacecraft encountered this comet in September, 2001.

Images at relative high airmasses and under pre-twilight conditions were
taken because of the unfavorable elongation. The night of the best
transparency conditions was selected for observations, although the 
seeing was larger than 2\arcsec.

%Fig. 2a.
\begin{figure*}
\begin{center}
\leavevmode
\caption{Comets 19P (left) and 29P (right). 
The images are rotated in a such way the solar direction is 
to the right.
The left subpanels show the observed images, while the right subpanels
are azimuthally renormalized images.
Graphs show radial sections of surface brightness profiles
(middle) in standard Gunn r and C$_2$. 
C$_2$ is plotted with respect to the zero magnitude of Gunn g.
Normalized intensity of spatial variations is shown in the bottom.
Local intensity ratio (lc) of spatial variations is expressed in percents
at positive and negative peaks.
Crosses refer to the radial section while solid line
shows the tangential section.
The images are 150\arcsec$\times$150\arcsec\ for 19P and 
100\arcsec$\times$100\arcsec\ for 29P.}
\end{center}
\label{f2a}
\end{figure*}

Although
the slope of the profile ($-$0.98) suggests an isotropic and
steady-state coma, the inner structure was quite complex. The nucleus
was far from the coma center, shifted to the antisolar direction. Dust
components formed a compact cloud in the inner 30\arcsec\ of the coma and
formed an impressive tail with forked structure. Gas components flowed
to the solar direction, with almost no gas observed in the tail. 
Surface photometry showed that the surface brightness of the
inner coma decreased faster in the antisolar direction (2 magnitudes in
7\farcs5 on the $r$ images) and slower toward the Sun (2 magnitudes in
11\arcsec). The outer coma became quite regular: 
it faded by 5 magnitudes 
in 53\arcsec\ to the antisolar direction and 37\arcsec\ to the solar
one.

As the azimuthally renormalized image shows, the behavior of the inner 
coma is
due to a jet-like outflow to the antisolar direction containing 16\% of the
total flux. This feature is detectable through 45\arcsec\ on the radial cross
section, which implies a proper length of 65$\cdot 10^3$ km 
on the assumption that
the jet is thin and lies on the solar radius. 
A comparison between the Gunn r and Comet C$_2$
profiles suggests that it consists solely of gaseous
components.
The tail is quite long and could be detected even beyond the
border of the image. Within a 125\arcsec\ distance to the nucleus,
it is brighter
than 24 mag/arcsec$^2$. Images and profiles are presented in 
Fig.\ 2/19P.
\bigskip\newline
\noindent
{\it 29P/Schwassmann--Wachmann 1}
\newline
\noindent
This unusual comet is well-known for its unpredictable outbursts (see Enzian
et al. 1997 and references therein). The nucleus seems to be perhaps the
largest one known in the Solar System, while its albedo is often estimated
to be over 15\% -- much higher than is measured for ``usual'' nuclei
(Jewitt 1990). Jets of the comet in outburst suggest rotational effects.
Several observers have tried to determine the rotation period: the
most recently published values are 14 and 32 hours (Meech et al., 1993).

We observed the comet during its 2001 outburst, which was detected
by Nakamura et al (2001) on May 17.69, 2001. 
During our observations and reductions, the main difficulty was
the crowded sky-field in the Milky Way. 
Our star-substraction procedure
removed about 3500 stars brighter than about 21 magnitudes from the
10\arcmin$\times$10\arcmin\ field.

Three months after the outburst the comet was 
12\fm68 in Gunn r, thanks to its position at opposition and high activity.
Due to the distance to the Earth, its coma was quite compact:
its surface brightness decreased 
2 magnitudes in 16\arcsec\ and 5 magnitudes in 35\arcsec.
The latter value corresponds to a diameter of 260$\cdot 10^3$ km.
On the  azimuthally renormalized, the well-known jet shows a spiral-like matter-rich 
and matter-poor part, with local contributions +37\% and $-$77\%, respectively.
The spinning shape is attributed to rotation of the nucleus. 

A ring-like structure, also visible with help of surface photometry, 
is not included in the non-radial part as it vanishes by subtracting the
azimuthal average from the coma image. The jet ends at 21\arcsec, while this
faint ring is suspected to be at 1\arcmin\ distance from the nucleus, having
22 mag/arcsec$^2$ surface brightness. Images are presented on the right side
of Fig.\ 2/29P.

We believe that the low counts in CN, CO+ and C$_2$ filters originate
from the continuum, and considering the estimated errors, only
higher limits of the emission rates are reported.
\bigskip\newline
\noindent
{\it C/2000 SV74 (LINEAR)}
\newline
\noindent
The comet, which reaches its perihelion on April 30.4, 2002,
was observed in order to monitor the pre-perihelion activity. 
At a solar distance of 4.244 AU, the measured brightness of
15\fm29 suggests that it will be quite bright near perihelion.

%Fig. 2b.
\setcounter{figure}{1}
\begin{figure*}
\begin{center}
\leavevmode
\caption{{\it (cont.)} Comets C/2000 SV74 (left) and C/2000 WM1 (right). 
Size of images presented is 75\arcsec$\times$75\arcsec\ for C/2000 SV74 and 
100\arcsec$\times$100\arcsec\ for C/2000 WM1.}
\end{center}
\label{f2b}
\end{figure*}

We measure a decrease in surface brightness of 
5 magnitude at 39\arcsec\
(114$\cdot10^3$ km) and 32\arcsec\ on the solar and antisolar sides,
respectively. No tail brighter than 24 mag/arcsec$^2$ was detected. 
The contribution of spatial variations was 33\% at the maximum of the non-radial
map and $-$22\% at the minimum. 
The simple structure of spatial structure is 
explained by a regular isotropic surface activity in the
presence of solar wind and radiation pressure. The slope of the coma profile
is $-1.21$, which seems to be shallower than profiles at larger solar
distances (see e.g. Szab\'o et al. 2001.). The feature could be explained
by strong and steady-state activity, containing mainly dust components.
This assumption of dust activity is also consistent with
the high $Af{\rho}$ (1892
cm) and the reddish color indices.
We detected weak emission in
C$_2$ and CO+.
\bigskip\newline
\noindent
{\it C/2000 WM1 (LINEAR)}
\newline
\noindent
The comet was observed 5 months before its perihelion date of January 22.68,
2002.  A compact circular coma was detected with a diameter of
$\sim20$\arcsec.

We also detected 
a thin tail, brighter than 24 mag/arcsec$^2$
within 90\arcsec\ and decreasing to about 25 mag/arcsec$^2$ at 120\arcsec\
(corresponding to 260$\cdot 10^3$ km projected length) from the
nucleus. 
Non-radial parts contribute 42\% (maximum peak) and
$-$60\% (minimum peak),
which we explain by 
radiation pressure on the coma. 
This is supported
by the slope parameter of $-$1.42, which is common for comets at similar
solar distance. 
The
apparently long and thin tail directed exactly in the antisolar direction 
is evidence that the Sun has the greatest influence on the coma and 
tail shapes.

The level of activity seemed to be lower than in the case of C/2000 SV74
($Af{\rho}$=275 cm): 
we measured relatively lower activity in C$_2$ and higher production of
CO+.  
This is in good agreement with the canonical view of
tail-formation, i.e. 
the dominant component of CO in cometary tails is observed 
at the examined wavelength (Arpigny 1995).
\bigskip\newline
\noindent
{\it C/2001 A2 (LINEAR)}
\newline
\noindent
The comet was in the very focus of the scientific interest during the summer
of 2001:
as of September, 2001, there have been 25 IAU Circulars issued
describing the evolution of this interesting comet. 
At the end of March, the
comet brightened 4 magnitudes in 4 days (Mattiazzo et al., 2001).
By the end of April, a double nucleus was detected (Hergenrother et
al., 2001). Further fragmentation was reported by Schuetz et al. (2001), 
and a CN
jet was reported by Woodney et al. (2001). At the second half of July, the
comet diminished 3 magnitudes and rapid light variations were observed;
these are explained by Kidger et al. (2001)
as the separation of small, short-lived splinters that may not
have been directly observable.

By the time of our observations, the total apparent magnitude decreased
below eleventh magnitude, and none of the multiple nuclei 
was detectable on our
images. Despite the ``calm'' behavior suggested by the run A2a (on 13th
August, 0:14 UT), a slight
increase of activity was detected on A2b images (on 15th August, 0:00
UT).
During the A2c run (on 16th August, 0:28 UT,
production rates of CN and C$_2$ increased by a factor of 4, while
dust production and $Af{\rho}$ did not vary much. 
The most surprising event between the nights is
the variation of the shape of C$_2$ profile. 
The
radiation pressure-dominated profile with narrow center turned into a
quite extended, symmetric profile with little central hump. Some parts of
the C$_2$ surface were brighter than the Gunn r continuum. That suggests a
similar outburst of activity in C$_2$ and CN as was reported by many
observers during the previous weeks (Kidger et al. 2001).

The Gunn r profile did not changed significantly between the two runs. 
In the
central 7\arcsec\, the surface brightness falls by 2 magnitudes, 
while on the
solar side it decreases by more than 5 magnitudes in about 70\arcsec\.
On the antisolar side the coma evolves into an impressively
bright and wide, V-shaped tail, which is  
brighter than 22 mag/arcsec$^2$ when it
leaves the field of view 
As the tail character suggests merely dust components, the
tail itself has also not been disturbed much by the observed small outburst. 
Altogether, the general morphology of C/2001 A2 is quite 
similar to the observed
behavior of 19P/Borrelly. 
An important difference is that in the case of C/2001
A2, neither type of jet has been detected, and
the azimuthally renormalized image shows a tail
rich in dust and detectable until the nucleus.

Although the contribution of spatial variations parts slightly decreased 
during
the outburst, the ratio of local contributions at minimum and maximum remained 
constant (i.e. 0.224/-0.217 vs. 0.159/-0.151) within an error of 4\%. That 
may be explained by a spherically symmetric outburst. 
In this case, 
the absolute values of non-radial parts do not vary,
but their contribution 
decreases as the absolute value of the radial part increases. 
Quantitatively, the radial part of the dust coma is increased by 39$\pm$2\%.
In this case, the brightness of the inner coma
should increase $-0\fm29$: that agrees with the observed value ($-0\fm24$) 
within the expected errors. 

A spherical outburst might
be caused by uniformly increasing activity on the whole 
surface of the nucleus, though the little fractions of active area commonly 
observed in comets seem to contradict this view.
Alternatively,
matter ejected in fans above the active area 
can be blended to globular shape if
we assume a fast-rotating nucleus. 
The assumption of a fast-rotating nucleus
agrees with the observed break-up of small fractions of the nucleus.

The measured 162--191 cm $Af{\rho}$ is smaller than 288 cm measured by
Schleicher (2001a), which is simply explained by the lower level of
activity. Images and profiles are shown in Fig. 2c.

%Fig. 2c.
\setcounter{figure}{1}
\begin{figure*}
\begin{center}
\leavevmode
\caption{{\it (cont.)} Comet C/2001 A2 on 13th (left) and 16th (right) August. 
Size of images presented is 150\arcsec$\times$150\arcsec. Note the 
strong change of the C$_2$ profile.}
\end{center}
\label{f2c}
\end{figure*}

\section{Discussion and concluding remarks}

Five comets were observed few days before new moon in August, 2001. The
comet producing the highest $Af{\rho}$ was 29P/Schwassmann--Wachmann 1 as it
was caught during its outburst. Other comets also showed unusually high
$Af{\rho}$ values, especially C/2000 SV74, which is predicted to become a
fairly bright, dust-rich comet at perihelion.

Inner comae of comets often show more matter ejected to the solar side
than elsewhere. This asymmetry of matter production is probably supported by
the warm solar side of the nucleus. 
Comparing the continuum-dominated
images and the emission-dominated ones, we find
a significantly changing coma composition for C/2001 A2 
in a smaller outburst. 
In the case
of 19P and C/2001 A2, the solar side is dominated by gaseous components,
due to interaction with the radiation pressure and solar
wind: dusty components are blown backwards and the solar side becomes
gas-rich. 
Outbursts, jets or proximity to the Sun may enrich 
gas in comae, and may lead to anomalous coma composition. 
We believe these factors explain the 
varying emission 
surface brightness profiles of C/2001 A2 in outburst.

%Table 5.
\begin{table}
\caption{Maxima and minima of spatial variations characterized as $s$ solar and
$a$ antisolar, $r$ radial or $t$ tangential-type peaks.
Distance from the nucleus is shown in $10^3$ km.} 
\label{peak}
\begin{center}
\begin{tabular} {lllrlrl}
\hline   
Ref. & R(AU) & $\Delta$(AU) & $D_{\rm max}$ & $lc_{\rm max}$ & $D_{\rm min}$ & $lc_{\rm min}$\\
\hline
19P & 1.400 &1.685 & 5.5 & 0.342$s$ & 3.7 & $-$0.398$t$ \\
29P & 5.919 & 5.121 & 26.1 & 0.370$s$ & 18.6 & $-$0.769$t$ \\
SV74 & 4.244 & 4.036 & 11.7 & 0.331$a$ & 5.9 & $-$0.218$rs$ \\
WM1 & 2.798 & 2.891 & 10.5 & 0.423$a$ & 7.3 & $-$0.603$rs$\\
A2a & 1.628 & 0.714 & 2.3 & 0.224$a$ & 1.3 & $-$0.217$rs$\\
A2c & 1.669 & 0.764 & 2.2 & 0.159$a$ & 1.4 & $-$0.151$rs$\\
\end{tabular}
\end{center}
\end{table}

In order to compare quantitatively the asymmetries in comets, the 
contributions of spatial variations are compared in Table\ 5.
Peaks of maximal intensity are distinguished by 
their location relative to the center of the coma:
$s$ means peaks on the solar side, $a$ denotes peaks on the 
anti-solar side.
We classify peaks of minima as follows:
$r$ denotes radial-type peaks
appearing near the radial section (solar or antisolar type), 
$t$ denotes
tangential-type peaks near the tangential section.
The local contribution of spatial variations at minimum peaks
seem to be well correlated with the geocentric 
radius. Similar but less
obvious correlations can be found for the maxima, too.
Generally, peaks of the azimuthally renormailzed images
develop farther from the nucleus,
and their contribution to the total coma intensity
increases with increasing solar distance. 
This conclusion is consistent with the 
view that solar wind and radiation pressure affects the cold and 
slowly outflowing coma of less active comets at a larger solar distance.

C/2000 SV74 seems to be an exception to the rule, as 
its coma
is significantly (by a factor of 3) less affected by spatial variations
than the other four comets. This character can be
hardly explained by the small phase angle
or the generally circular appearance, 
and supports the idea that high level matter production is present that can
challenge the solar wind.

Dependencies between Gunn-colors and  $Af{\rho}$
parameter have been found for comets. Below we summarize the correlation,
their standard errors (in brackets) and their regression coefficient. 
Independent correlations with better regression coefficient than 
0.80 or $-$0.80 have been accepted.

\begin{tabular} {ll}
\\
$(v-g)=-0.51(20)+0.214(7)\cdot {\rm log}~Af{\rho}[cm]$ & \\
\hskip1cm regr. coeff.=0.87 &\\
$(r-j)=0.12(5)+0.7(3)\cdot (v-g)$ & \\
\hskip1cm regr. coeff.=0.81 & \\
$(j-z)=0.19(4)-0.8(2)\cdot (r-j)$ & \\
\hskip1cm regr. coeff.=-0.88 &\\
\\
\end{tabular}

We explain the correlation between $v-g$ color and 
$Af{\rho}$ as a scattering effect of dust.
Matter particles scatter the short-wavelength violet color the most, thus,
the more dust present in the coma, the more "reddish" $(v-g)$ its color.
Continuum colors of 29P/Schwassmann-Wachmann 1 are measured to be slightly less
reddish than previously published values: $B-V=0.8$
(Hartmann et al., 1982), $V-R=0.502$ and
$R-I=0.492$ (Meech et al. 1993), which may be transformed into the Gunn-system
with help of the transformations determined by Kent (1985) and J\o rgensen
(1996), yielding $v-g=0.478$, $g-r=0\fm028$.
Meech et al. (1993) attribute
reddish colors to a large distance from the Sun.
In present paper, the experimental correlation between $Af{\rho}$ and
(v-g) also explains quite reddish colors.

Correlation between $r-j$ and $j-z$ may be an artificial effect. 
Examining the individual comets, the $j-z$ color is found to 
correlate with solar distance as the radiation 
contribution in the near-infrared has
increasing effect with decreasing solar distance.

In order to understand production rates,
we have
compared our data to the extensive set of A95. In the case of 19P/Borrelly,
production rate ratios are in a perfect agreement with the previous
results, as mentioned in A95. 
The detected lance-head shape of the
coma is similar to the HST images taken by Lamy et al. (1998). 
Their slope
parameter varying around $-$1 with moderate angular variations is in a perfect
agreement with our observations.
We note that Schleicher (2001b) published
$Af{\rho}$ measurements in September, which were half those
measured by us.  The slope parameter $G=-1.9$ of Schleicher (2001b) is
also not comparable with our smooth profile.

C/2000 SV74 and C/2000 WM1 have similar appearances, although during
their further evolution significant structural differences may develop,
due to the difference in their perihelion distances (3.54 and 0.56 AU).
The abundance of C$_2$ in C/2000 SV74 is much higher than usual:
one can find only 3 comets (P/Russel 4,
C/Shoemaker 1984 XII, C/Shoemaker 1984 XV) with higher C$_2$/CN production
among the 85 objects discussed in A95.

C/2001 A2 showed a significant change in composition during its outburst.
Compared to the statistics and Fig.\ 4c of A95, the [$Af{\rho}$]-[CN] seems to
be anomalously high in ``quiescent'' state ($-22.07$), while is also a bit
high, but not peculiar, in ``outburst'' ($-22.74$);
these values fall within the
$-23.1\pm0.9$ interval for all of the comets discussed in A95. As a possible
explanation, one can imagine that C/2001 A2 was in a period of
anomalously low
activity, a ``negative outburst'', during the time of high
[$Af{\rho}$]-[CN] production (e.g. A2a, on 13th August, 0:14 UT).
The normal behaviour in
outburst-like events is to show increased gas activity. 

Comets 19P and C/2001 A2 show similarly bluish colors and CN-rich gas
production. Low [C$_2$]-[CN] rates allow us to classify C/2001 A2 as a
Borrelly-type comet (Fink et al. 1999).
Note that the drastic variations observed
during some months previously might also influence the
classification.

Finally, we summarize our results as follows.

\noindent 1. The use of Gunn photometric system in
structural and production analysis of comets is demonstrated. 
When augmented with comet and continuum interference filters,
the system combines the advantages of the two systems
usually preferred 
in cometary astronomy (Johnson filters for morphology and 
narrow-band filters for spectrophotometric studies).

\noindent 2. A technique for generating and analyzing azimuthally
renormalized images or non-radial residual
maps is discussed, and the power of this tool to study morphology is
demonstrated. In the case of C/2000 SV74, C/2000 WM1 and C/2001 A2, simple
comae with slight solar formation effects are observed. In the case of
19P/Borrelly, slope parameter was almost
$-1$, but spatial variations were observed to make 
moderate contributions.
29P/Schwassmann-Wachmann 1 was observed in outburst, and a well-developed jet
and a ring at larger nuclear distance were detected.
Variations from spherical outflow are indicated
by local contribution parameters at the peaks of the
azimuthally renormalized images. Generally, the contribution
of the non-radial parts 
to the whole coma was found to increase with the solar distance.

\noindent 3.
C/2000 SV74 and C/2000 WM1 were observed a few months before their
perihelion.  Large $Af{\rho}$ values, the extended coma of C/2000 SV74 
and the nice tail of C/2000 WM1 
suggest that they will be interesting objects in perihelion. 
C/2000 SV74
showed unusually high [C$_2$]-[CN] production ratio,
and developed a nearly spherical coma at a distance of 4 AU.
 During our
observations, C/2001 A2 suffered a smaller outburst, ejecting mainly
gaseous components.  The ejected matter evolved in a spherically
symmetric manner.
The amount
of gas increased by a factor of 4 during the outburst, while 
the dust components increased by about 39\%. 
Its matter production is quite atypical when low level of activity is present.

\begin{acknowledgements}

This research was supported by FKFP Grant 0010/2001,
Pro Renovanda Cultura Hungariae Grant DT 2001. m\'ajus/21.,
OTKA Grants \#T032258 and \#T034615, ``Bolyai J\'anos'' Research
Scholarship of LLK from the Hungarian Academy of Sciences, ``Koch S\'andor''
Foundation (ref. number KSA 2001/8) and Szeged Observatory Foundation.
The warm hospitality and helpful assistance 
of the staff of Calar Alto Observatory is gratefully acknowledged.
The authors also acknowledge suggestions and careful reading of the manuscript
by Dr. Michael Richmond.
The NASA ADS Abstract Service was used to access data and references.

\end{acknowledgements}

\end{document}